\documentclass[conference]{IEEEtran}
\IEEEoverridecommandlockouts

\usepackage{cite}
\usepackage{color,soul}
\usepackage{amsmath,amssymb,amsfonts}
\usepackage{graphicx}
\usepackage{textcomp}
\usepackage{xcolor}
\usepackage{braket}
\usepackage{algorithm}
\usepackage{algpseudocode}
\usepackage[a4paper, total={184mm,239mm}]{geometry}

\def\BibTeX{{\rm B\kern-.05em{\sc i\kern-.025em b}\kern-.08em
    T\kern-.1667em\lower.7ex\hbox{E}\kern-.125emX}}

\title{Forensics of Error Rates of Quantum Hardware}

\author{
\IEEEauthorblockN{Rupshali Roy}
\IEEEauthorblockA{
    School of EECS\\
    Penn State University, PA, USA\\
    rzr5509@psu.edu
}
\and
\IEEEauthorblockN{Swaroop Ghosh}
\IEEEauthorblockA{
    School of EECS\\
    Penn State University, PA, USA\\
    szg212@psu.edu
}
}

\begin{document}

\maketitle

\begin{abstract}
There has been a rise in third-party cloud providers offering quantum hardware as a service to improve performance at lower cost. The qubit technologies, basis gate set, noise behavior, speed and coupling architecture are among the various factors that differ among various backends. Although these providers provide flexibility to the users to choose from several qubit technologies, quantum hardware, and coupling maps; the actual execution of the program is not clearly visible to the customer. 
The success of the user program, in addition to various other metadata such as cost, performance, \& number of iterations to converge, depends on the error rate of the backend used. Moreover, the third-party provider and/or tools (e.g., hardware allocator and mapper) may hold insider/outsider adversarial agents to conserve resources and maximize profit by running the quantum circuits on error-prone hardware. Thus it is important to gain visibility of the backend from various perspectives of the computing process e.g., execution, transpilation and outcomes. In this paper, we estimate the error rate of the backend from the original and transpiled circuit. Although many quantum services providers publish the error rates of their backends, we assume that such information may not be accurate and/or correspond to the actual hardware allocated to the program.
For the forensics, we exploit the fact that qubit mapping and routing steps of the transpilation process select qubits and qubit pairs with less single qubit and two-qubit gate errors to minimize overall error accumulation, thereby, giving us clues about the error rates of the various parts of the backend. We ranked qubit links into bins based on ECR error rates publicly available, and compared it to the rankings derived from our investigation of the relative frequency of a qubit link being chosen by the transpiler. For upto 83.5\% of the qubit links in IBM Sherbrooke and 80\% in IBM Brisbane, 127 qubit IBM backends, we are able to assign a bin rank which has a difference upto 2 with the bin rank assigned on the basis of actual error rate information. 
\end{abstract}


\section{Introduction} \label{intro}
Cloud based quantum computing services have greatly catalyzed the development of quantum computers. Quantum hardware is becoming more and more accessible in the Noisy Intermediate-Scale Quantum (NISQ) era, enabling research \& development at lower expenses. Multi-qubit processors with different layouts, architectures, and qubit technologies are available through cloud providers. Therefore, designers get a wide variety of choices to work with.

The hardware available through such providers is black-boxed. There may be multiple choices of coupling maps that are isomorphic in terms of noise, architecture, and execution speed) within a backend or among a suite of backends. The program could be mapped onto any of these even though the backend, the coupling map, and qubits are specified by the user. Hardware allocation policies choose a coupling map that is a subgraph of the backend structure, to minimize SWAP operations, circuit depth, gate count and aggregate error rate through several transpiler passes. But the user has no way to know if the mapping with the lowest possible error was chosen by the backend to transpile the circuit. The cloud vendor may not consider this as a gross violation of trust but this has the potential to significantly jeopardize the performance, execution speed \& optimization quality. In a multitenant environment it becomes even more challenging where the same hardware is shared by several users/programs. Community-based partitioning \cite{b01} of hardware is used to assign physical qubits to the queued user programs. Malevolent jobs that could inject fault into the victim's program through crosstalk are identified by metric-based measurements. These are then restricted from execution. However, it is not customary to adopt such security-aware attitudes while developing regular allocators \& schedulers. Moreover, it is still possible that other programs sharing the hardware still manage to divert the program to be allocated to more error-prone qubits, in spite of such security measures. The tools used for implementing scheduling, compilation and allocation policies and metrics are proprietary, thus making it impossible to validate if the decisions made by the tool corresponds to the highest fidelity execution requirements or not. 

To cut down on time and resources, third party tools, which may not be as trustworthy, are being adopted. This exacerbates the issue of misrouting/misallocation of programs to error prone physical qubits. 
This problem will be further amplified by the usage of third party quantum cloud vendors, software and/or hardware that is not from a trusted source, since the integrity of the toolchain is also questionable.

An adversary could be motivated to execute the circuit on more error-prone qubits while charging the user for higher performing qubits, to conserve expensive quantum resources, increase profit margin, degrade the computation quality thus causing denial-of-service, and to inject failures. Thus we need forensics which is similar to reverse engineering to gain insight into how the cloud service provider takes decisions through deliverables/deliverable characteristics that we get from it, such as transpiled programs, speed of execution, time to converge et cetera. This would help us validate vendor claims, learn more about the decision making policies \& algorithms followed by the provider, thus making the process trustworthy. Successful forensic investigation would allow customers to (i)\emph{Fingerprinting of hardware through identifying characteristics:} It will be possible to identify the hardware through its characteristics such as qubit quality characterisation, and times for gate execution where the job was executed. (b) \emph{Creating trust in quantum services:} The user would be able to validate if the vendor honored their claims such as, allocation of high fidelity qubits for computation establishing trust in the process. (c) \emph{Locating bugs and/or loopholes in toolstack:} It will enable detection of bugs in the toolchain and find loopholes such as parallel job allocation that causes injection of crosstalk-induced faults among the jobs.


\textbf{Contributions} In this paper, we present an elementary step in forensics of quantum systems by characterizing the backends in terms of the error rates of its constituent qubit links on the basis of performance. To achieve this objective, we extract the physical qubit topologies from a number of random transpiled copies of utility programs using the forensic implementation from \cite{b5}. \footnote{Usually, quantum hardware providers make the backend error information public, but we assume that the circuit may not have been executed on the higher fidelity qubits but on the more error-prone ones instead.} The qubit edges for the backend are then ranked according to frequency of occurrence and classified based on performance, thus characterizing the backend with this information. We assume that the user submits the program to the quantum service provider for transpilation on the specified backend according to its allocation policy, then the provider sends back the transpiled program along with the execution results. However, the user is unable to validate whether the cloud vendor honored their claim to execute the circuit on the physical qubits that would give us the highest possible fidelity. By characterizing the qubits into performance-ranked bins and comparing this analysis with publicly available information on the target backend, we can bridge the above-mentioned opacity.

In the remainder of this paper, we provide background on quantum computing and related works in Section II. The methodology and results are presented in Section III. Conclusions are drawn in Section IV.

\section{Background} \label{background}
\subsection{Preliminaries}
\subsubsection{Qubits} The fundamental unit of quantum information is a qubit. It is usually represented by a binary system, having basis states denoted as $\ket{0}$ and $\ket{1}$. Unlike classical bits, which can only be in one state (0 or 1) at a time, qubits can exist in a superposition of these two states. The general state of a single qubit is described by a linear combination $\psi$ = $\alpha$ $\ket{0}$ + $\beta$ $\ket{1}$, where $\alpha$ and $\beta$ are complex coefficients. The squares of the magnitudes of $\alpha$ and $\beta$ give the probabilities of measuring the qubit in the $\ket{0}$ or $\ket{1}$ state, respectively, with the sum of these magnitudes equal to 1.

\subsubsection{Quantum gates} These are operations applied on single or multiple qubits (using microwave or laser pulses for superconducting qubits), to transform the qubit state, entangle qubits, or generate superpositions. They are mathematically expressed in the form of unitary matrices, satisfying the condition $\mathbf{U}^\intercal U = I$, where $\mathbf{U}^\intercal$ is the conjugate transpose of $U$, and $I$ is the identity matrix. Some common quantum gates include the Hadamard gate, the rotation gates (Rx, Ry, Rz), the Pauli-X gate, and the controlled-NOT (CNOT) gate. The Hadamard gate creates an equal superposition of the $\ket{0}$ and $\ket{1}$ states. Rotation gates Rx (or Ry) rotate the state vector of the qubit around the x-axis (or y-axis) of the Bloch sphere by a phase given in radians. The Pauli-X gate flips the qubit's state. The CNOT gate is a two-qubit gate that flips the target qubit if the control qubit is in the $\ket{1}$ state.  
\subsubsection{Basis gates and coupling constraints} Only a specific set of single- and multi-qubit gates are supported by quantum computers. This is known as the set of basis gates or native gates of the hardware. IBM quantum devices, for instance, use native gates like the CNOT (two-qubit), u1, u2, u3, and id (single-qubit) gates. If there are non-native gates (e.g., a Toffoli gate) in the quantum circuit, these must be decomposed into basis gates before execution. In addition, two-qubit operations like CNOT are only allowed between physically connected qubits, a restriction referred to as a coupling constraint. SWAP gates are inserted to satisfy these constraints if a two-qubit gate involves unconnected qubits.

\subsubsection{Error rates} NISQ-era quantum computers are affected by noise from a variety of sources \cite{b6}. In order to maintain their state, qubits need to be perfectly isolated from the environment. Of course in a real scenario that is impossible to achieve and the inevitable interaction, or "measurement" of sorts, causes information to leave the system. Over time, the qubit collapses to a classical state due to leakage and/or relaxation. The time for which the qubit can maintain its state is called its coherence time. This is usually reported via \(T_1\) time which measures the time it takes to collapse to the $\ket{0}$ from $\ket{1}$ state. This measures the expected loss of energy from the system. Another metric used is \(T_2\) or dephasing time that measures the time to collapse from the superposition state to either $\ket{0}$ or $\ket{1}$.

This problem of noise gets exacerbated when we apply any operation on a qubit. Imperfection in quantum operations leads to incorrect quantum states, due to over or under-rotation owing to imperfect calibration. These errors are called gate errors that are reported for both single and 2 qubit gates, specifically for the native gates of the hardware in question. Some examples of single qubit gate errors would be Pauli-X error, ID error etc in IBM backends. For 2 qubit gates, ECR errors are reported.

Another source of error that is difficult to characterize is the crosstalk due to undesired interaction between qubits. Such undesirable interaction may cause mixing of quantum states of the involved qubits, or decoherence.

\subsubsection{Compilation}  
\begin{figure*} 
    \centering
        \vspace{-3mm}
        \includegraphics[width=\linewidth]{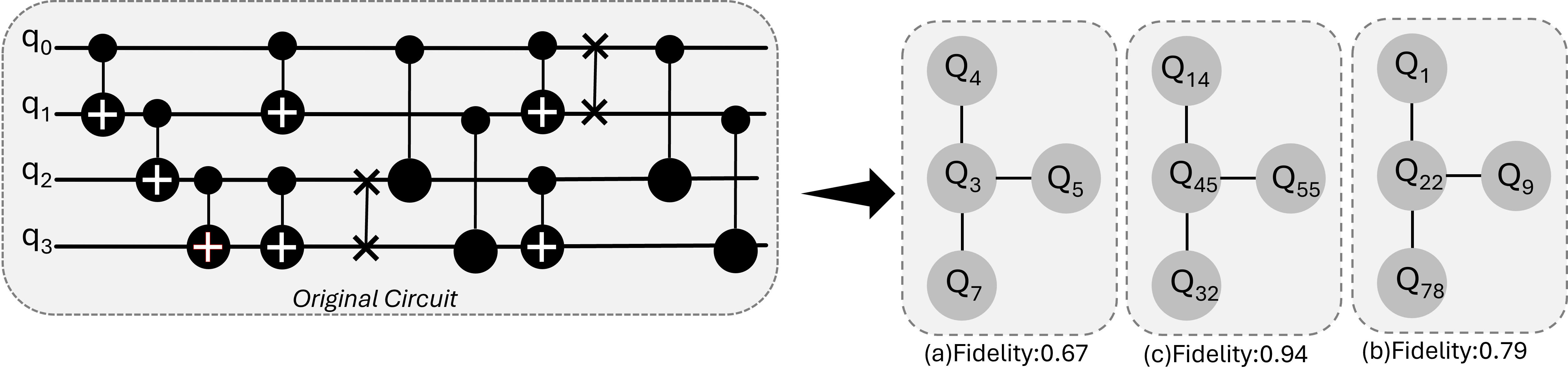}
        \vspace{0mm}
        \caption{Optimized mapping choices available for a transpiler. The mapping having highest fidelity is chosen, (c) in this case.}
        \label{choice_map}
\end{figure*}

Quantum circuit compilers transform user circuits to comply with the coupling constraints of the hardware, often inserting SWAP gates to achieve this. Circuit optimization also involves merging, canceling, or reordering single- and multi-qubit gates, as well as performing rotations. IBM Qiskit provides support for barriers, which prevent optimizations across specific sections of the circuit.The compilation process optimizes the circuit to have least possible aggregate gate error, gate count and circuit depth. In other words, the choice of qubit links made by the compiler for a given circuit transpilation process also qualitatively indicates which qubit links have lower error rate and are thus better performing. For example, in Fig. \ref{choice_map} we see 3 different potential mapping choices available to the transpiler after optimizing for gate count, circuit depth and number of SWAP gates. Since choice (c) has the highest fidelity among the three, it is selected as the final mapping for the example circuit. 

\subsection{Related works}
The existing work on quantum computing forensics involves extraction of the physical topologies of transpiled circuits \cite{b5}. Two qubit operations are only allowed between qubits that are physically connected on the coupling map of the backend, and SWAP gates are inserted to execute two-qubit gate operation between unconnected qubits. Leveraging this feature of transpilers, a heuristic is proposed by the authors to extract the coupling map from the transpiled circuit. However, this approach does not reveal the error rates of the backend qubits. There has been work on fingerprinting of quantum hardware where a probing circuit is used to capture the unique error characteristics of quantum devices \cite{b1}. The results of the execution of the probing circuit act as a device-side fingerprint of the quantum hardware when the user inspects the service. In \cite{b}, a power side channel-based attack is proposed that retrieves information about the qubit mapping of the circuit. However, we focus on forensics of the 2-qubit error rates. In addition, the power consumption of the quantum hardware is privileged information and is not available to the users.
The quantum servers are fingerprinted by running a user's circuit with two different levels of noise in \cite{b2}, utilizing the resultant performance gap as a fingerprint. In \cite{b3}, the frequency of qubits is used to identify quantum computers based on transmon qubits by noting that the frequencies of individual qubits are unique. This is mainly due to process variations during manufacturing, creating distinct physical properties for each qubit. A quantum physically unclonable function (QuPUF) \cite{b4} fingerprints the requested hardware on the transpiled circuit. Two flavors of QuPUF circuits are proposed based on superposition and decoherence. The fingerprinting techniques, however, rely on circuits without any real functionality. Hence, they are not relevant for forensics purposes. 

\section{Methodology and Results}
\begin{figure*}
        \centering         
        \begin{minipage}{0.5\textwidth}
                \centering
                    \includegraphics[width=\columnwidth]{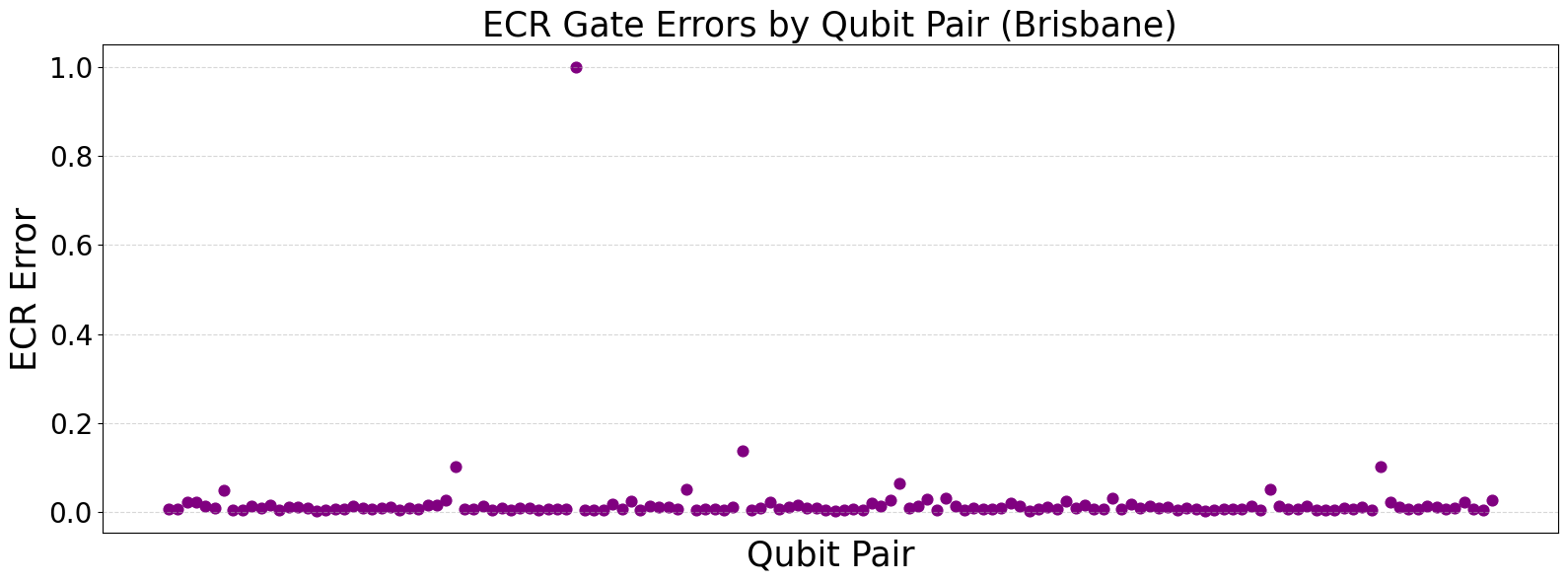}
                 \vspace{0mm}
                 (a)
        \label{brisbane_ECR}
        \end{minipage}%
        \begin{minipage}{0.5\textwidth}
                \centering
                    \includegraphics[width=\columnwidth]{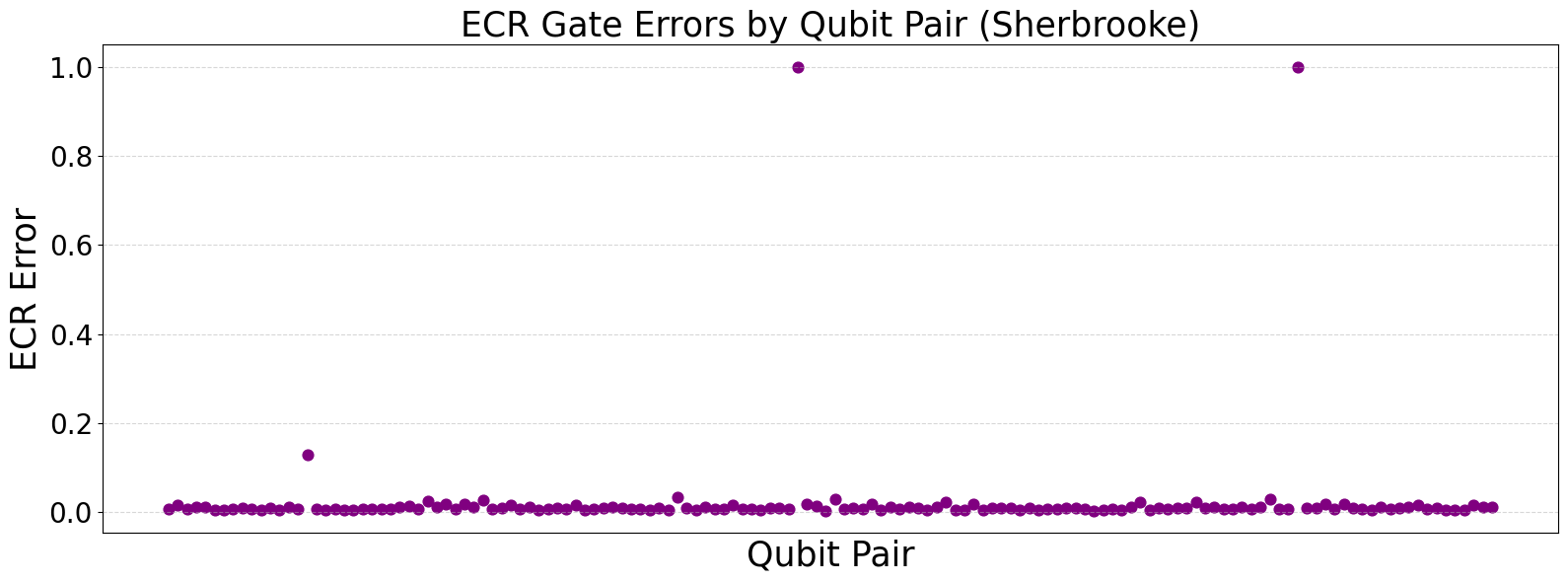}
                 \vspace{0mm}
                 (b)
        \label{sherbrooke_ECR}
        \end{minipage}%
        \vspace{0mm}
        \caption{ECR error rate information publicly available for IBM Brisbane and IBM Sherbrooke.}  
        \label{ecr}
\end{figure*}

\begin{figure*}
        \centering         
        \begin{minipage}{0.5\textwidth}
                \centering
                    \includegraphics[width=\columnwidth]{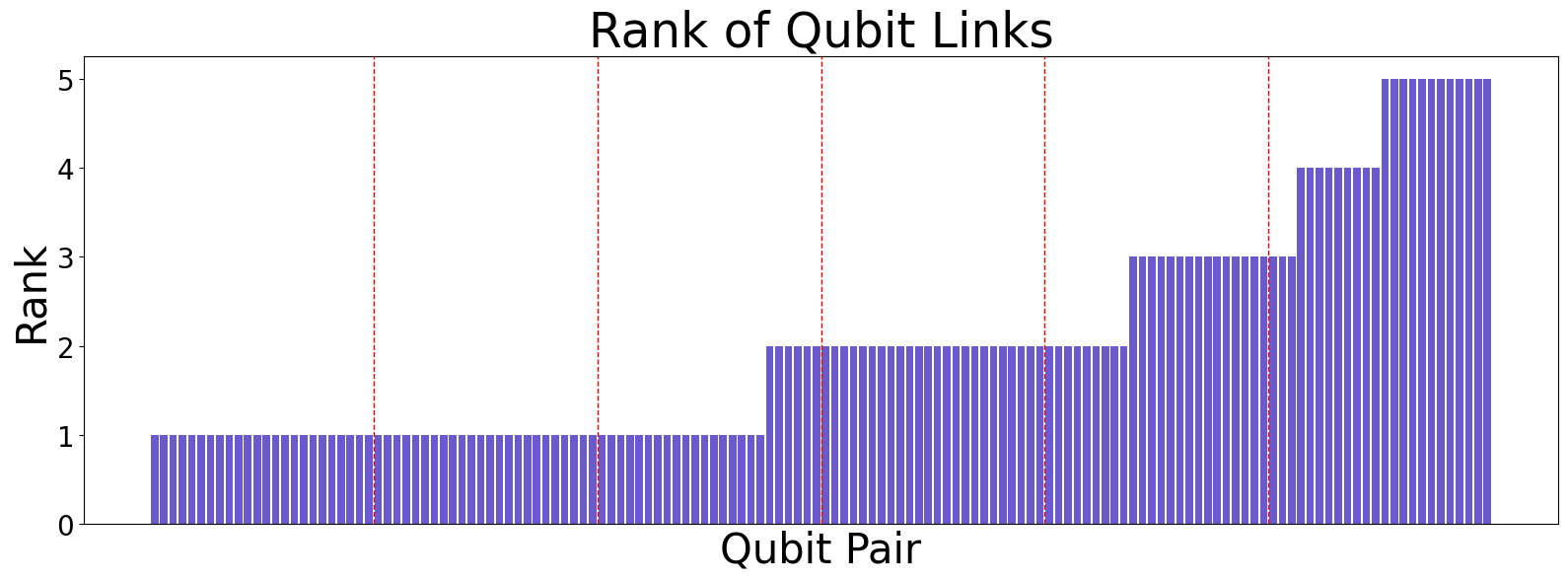}
                 \vspace{0mm}
                 (a)
         \label{brisbane}
        \end{minipage}%
        \begin{minipage}{0.5\textwidth}
                \centering
                    \includegraphics[width=\columnwidth]{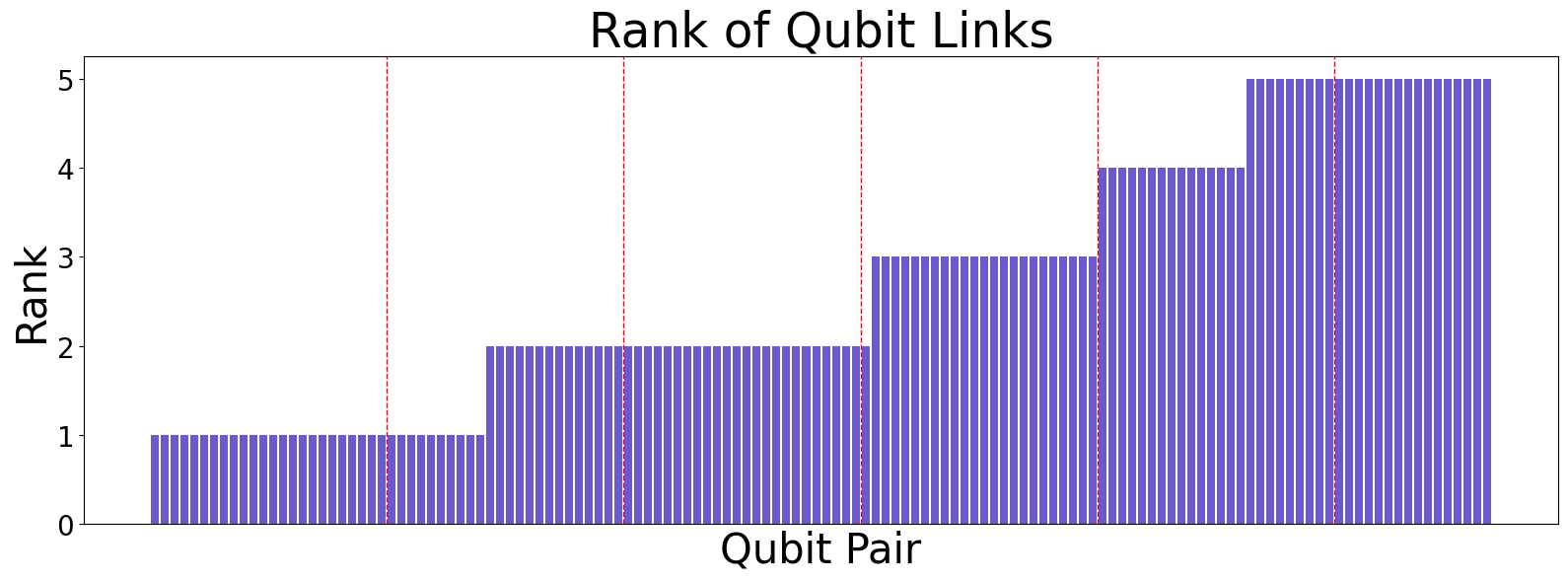}
                 \vspace{0mm}
                 (b)
        \label{sherbrooke}
        \end{minipage}%
        \vspace{-1mm}
        \caption{Ranking of the qubit links on the basis of our analysis for (a) IBM Brisbane (b) IBM Sherbrooke.}  
        \label{ranking}
\end{figure*}

\begin{figure*}
        \centering         
        \begin{minipage}{0.5\textwidth}
                \centering
                    \includegraphics[width=\columnwidth]{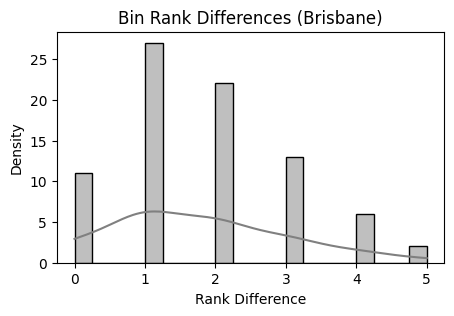}
                 \vspace{0mm}
                 (a)
         \label{brisbane_new}
        \end{minipage}%
        \begin{minipage}{0.5\textwidth}
                \centering
                    \includegraphics[width=\columnwidth]{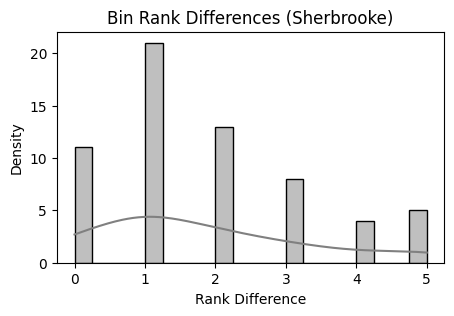}
                 \vspace{0mm}
                 (b)
        \label{sherbrooke_new}
        \end{minipage}%
        \vspace{-1mm}
        \caption{Density distribution of the differences between bin rank derived from our analysis and the publicly available error rate information for (a) IBM Brisbane (b) IBM Sherbrooke.}  
        \label{distribution}
\end{figure*}

\subsection{Error Rate Extraction Methodology}
It is noted that when a circuit is transpiled and mapped to a physical hardware topology, the transpiler decomposes the circuit to the hardware's native gates and then finds an optimized mapping through several passes. When the logical qubits involved in multi-qubit operations are mapped onto the physical qubits that may not be connected to each other in the backend structure, SWAP operations are used to ensure the operations can still be carried out. We also note that the transpiled QASM code lists the physical qubit numbers of the backend where the logical qubit numbers have been mapped. Leveraging this and the presence of these SWAP gates between disconnected physical qubits to ensure logical connectivity, \cite{b5} shows how the physical qubit topology can be extracted from a transpiled circuit. Several possible logical-to-physical mappings are found that have the lowest possible gate count, circuit depth and least number of SWAP gates. Now the transpiler calculates which of these has the highest fidelity using the following metric: 
\[E = (1 - E_{g_1})^{n_1}*(1 - E_{g_2})^{n_2}*....(1 - E_{g_N})^{n_N}\]
where \(E_{g_i}\) is the error rate for each gate type applied on a specific qubit pair/single qubit in the circuit, and \(n_i\) is the count for that gate. Hence, the higher fidelity the chosen qubit link has, the greater the overall fidelity of the chosen mapping. Therefore, the transpiler attempts to choose higher fidelity qubit links so as to minimize the aggregate error. 

Our approach is to extract the relative error rate of the qubit links from a suite of transpiled circuits. We note that the highest quality links are likely to be selected during transpilation for many circuits whereas the poor quality links are least likely to be selected for any circuit. Therefore, we can count the number of times each link is selected which will roughly indicate the relative error of the links compared to each other. Instead of absolute ranking of qubit links we create error rank bins. The links within the same bin have similar error rates. For example, the links in the first bin are selected most of the time, hence they are expected to be of best quality whereas the qubit links in the second ranked bin have second best fidelity and so on.

We then compare this bin ranking with the bin ranking based on the real error information for the chosen backend that is publicly available. 

\subsection{Results}
For our analysis we use 5 random 100-qubit quantum circuits generated using the random\_circuit module in Qiskit. These circuits are transpiled on the open-access real IBM hardware, namely IBM Sherbrooke and IBM Brisbane. These are 127 qubit Eagle R3 processors. For this forensic method to be considered valid, we consider the scenario that several jobs are submitted to the cloud vendor with target hardware. That is how we have access to multiple transpiled programs that will aid our forensic investigation. Our circuits are large enough to allow us to use all the qubit connections in the backends. Using the procedure followed in \cite{b5}, we extract the physical topologies of the circuits. 

We run the simulations using IBM Qiskit locally on an Intel(R) Core(TM) 7 150U CPU with Intel Graphics (1.80 GHz) machine with 16 GB RAM. We create 6 equal-sized bins for the 144 qubit edges in each backend on the basis of the frequency of occurence (Fig, \ref{ranking}). Each bin in this case holds 24 qubit links. The 24 most frequntly chosen edges are held in the first bin, then the next 24 most frequently chosen edges are put in the second bin and so on and so forth. We then create a similar bin ranking for the qubit edges based on the ECR (2 qubit) error information publicly available on the IBM website for these backends (Fig. \ref{ecr}). We then observe the difference in the bin rank from our analysis and the bin rank from the actual error information. We see that we are able to predict the bin rank within a good range of accuracy for a major portion of the backends. For IBM Brisbane, we predicted the bin rank accurately or with a difference of upto 2 for more than 80\% of the qubit edges (Fig. \ref{distribution}(a)). For the IBM Sherbrooke backend, around 83.5\% of the qubit edges are predicted within bin rank accurately or a difference of upto 2 (Fig. \ref{distribution}(b)). 

\section{Conclusion}
Quantum hardware varies with respect to basis gate set, noise behavior, coupling architecture and speed, among other parameters. There are several choices of quantum hardware, qubit technologies, and coupling maps available to the user. But the circuit execution is a black-box operation in the quantum cloud. The choice of physical qubits used for execution plays a major part in determining the fidelity of the user program. However, the third-party provider may be untrustworthy and/or the hardware scheduler may be buggy. Therefore, the provider may run the quantum circuits on more error-prone and/or unreliable hardware. Hence, there is a need to gain transparency using forensics on the backend which will help establish trust in the quantum cloud services and validate the qubit quality allocated during transpilation. In this work we characterize a backend on the basis of the fidelity of its constituent qubit edges. We group the edges into bins based on the frequency of being chosen by the transpiler in a suite of circuits transpiled on that backend. We compare our findings against a bin ranking derived from the error rate information publicly available demonstrating promising results.

\vspace{12pt}


\begin{thebibliography}{00}
\bibitem{b0} https://quantum-computing.ibm.com/services/resources/docs/resources
/manage/systems/queue
\bibitem{b01}Upadhyay, Suryansh, and Swaroop Ghosh. "SHARE: Secure Hardware Allocation and Resource Efficiency in Quantum Systems." arXiv preprint arXiv:2405.00863 (2024).
\bibitem{b1} Wu, Jindi, Tianjie Hu, and Qun Li. "Detecting Fraudulent Services on Quantum Cloud Platforms via Dynamic Fingerprinting." arXiv preprint arXiv:2408.11203 (2024).
\bibitem{b}Chuanqi Xu, Ferhat Erata, and Jakub Szefer. 2023. Exploration of Power Side-Channel Vulnerabilities in Quantum Computer Controllers. In Proceedings of the 2023 ACM SIGSAC Conference on Computer and Communications Security (CCS ’23). Association for Computing Machinery, New York, NY, USA, 579–593. https://doi.org/10.1145/3576915.3623118
\bibitem{b2} Wu, Jindi, Tianjie Hu, and Qun Li. "Q-ID: Lightweight Quantum Network Server Identification through Fingerprinting." IEEE Network (2024).
\bibitem{b3} Smith, Kaitlin N., Joshua Viszlai, Lennart Maximilian Seifert, Jonathan M. Baker, Jakub Szefer, and Frederic T. Chong. "Fast fingerprinting of cloud-based nisq quantum computers." In 2023 IEEE International Symposium on Hardware Oriented Security and Trust (HOST), pp. 1-12. IEEE, 2023.
\bibitem{b4}K. Phalak, A. A. -. Saki, M. Alam, R. O. Topaloglu and S. Ghosh, "Quantum PUF for Security and Trust in Quantum Computing," in IEEE Journal on Emerging and Selected Topics in Circuits and Systems, vol. 11, no. 2, pp. 333-342, June 2021, doi: 10.1109/JETCAS.2021.3077024.
\bibitem{b5}Roy, Rupshali, Archisman Ghosh, and Swaroop Ghosh. "Forensics of Transpiled Quantum Circuits." arXiv preprint arXiv:2412.18939 (2024).
\bibitem{b6}Salonik Resch and Ulya R. Karpuzcu. 2021. Benchmarking Quantum Computers and the Impact of Quantum Noise. ACM Comput. Surv. 54, 7, Article 142 (September 2022), 35 pages. https://doi.org/10.1145/3464420

\end{thebibliography}
\end{document}